\documentclass[doublecol]{epl2}
\usepackage{color}

\title{Controlling the spontaneous spiking regularity via channel \\ blocking on Newman-Watts networks of Hodgkin-Huxley neurons}
\shorttitle{Controlling the spontaneous spiking regularity via channel blocking...}

\author{Mahmut Ozer,\inst{1} Matja{\v z} Perc,\inst{2} Muhammet Uzuntarla\inst{1}}
\shortauthor{Mahmut Ozer \textit{et al.}}

\institute{
\inst{1}Zonguldak Karaelmas University, Engineering Faculty, Department of Electrical and Electronics Engineering, 67100 Zonguldak, Turkey\\
\inst{2}Department of Physics, Faculty of Natural Sciences and Mathematics, University of Maribor, Koro{\v s}ka  cesta 160, SI-2000 Maribor, Slovenia}


\abstract{We investigate the regularity of spontaneous spiking activity on Newman-Watts small-world networks consisting of biophysically realistic Hodgkin-Huxley neurons with a tunable intensity of intrinsic noise and fraction of blocked voltage-gated sodium and potassium ion channels embedded in neuronal membranes. We show that there exists an optimal fraction of shortcut links between physically distant neurons, as well as an optimal intensity of intrinsic noise, which warrant an optimally ordered spontaneous spiking activity. This doubly coherence resonance-like phenomenon depends significantly, and can be controlled via the fraction of closed sodium and potassium ion channels, whereby the impacts can be understood via the analysis of the firing rate function as well as the deterministic system dynamics. Potential biological implications of our findings for information propagation across neural networks are also discussed.}

\begin{document}

\maketitle

\section {Introduction}
Information processing within the nervous system involves noisy components \cite{r1}. One major source of noise within neurons is due to the stochastic dynamics of voltage-gated ion channels embedded in neuronal membranes \cite{r2}. Voltage-gated ion channels are involved in generating and propagating electrical signals through neuronal membranes. Hodgkin and Huxley (HH) \cite{r3} first introduced conductance-based models of these channels. However, the original HH model does not take into account the stochastic dynamics of voltage-gated channels, but provides a deterministic description of the membrane potential, which is valid only within the limit of very large cell sizes. When the population of ion channels is finite, the stochastic dynamics of voltage-gated ion channels (or ion channel noise) causes subthreshold fluctuations in the membrane voltage \cite{r2}, and can have significant impacts on the neuronal dynamics \cite{r5,r6,r7,r8,r9,r10,r11,r12}.

The intensity of the channel noise is related to the number of ion channels, but its actual impact is determined by the number of channels that are open near the threshold for spike firing. Schneidman \textit{et al.} \cite{r7} showed that there is a short distance in terms of the number of open channels between firing and non-firing stable states, and that fluctuations due to only a few channels are responsible for the transition between these two stable states. They also reported that this spontaneous transition is the cause for the missing spikes, the subthreshold membrane voltage, as well as spontaneous spikes. Therefore, controlling the number of working ion channels for a given membrane patch is of great importance, particularly to understand the impact of a specific ion channel type on neuronal dynamics. In this context, some toxins such as tetraethylammonium (TEA), tetradotoxin (TTX) and saxitoxin (STX) are widely used in experiments to block or reduce the number of specific ion channels \cite{r13}. It is also possible to examine the effects of changing the number of specific ion channels on neuronal dynamics through computational models. Schmid \textit{et al.} \cite{r14,r15} investigated the regularity of spontaneous spiking activity of a stochastic HH model at a single neuron level, and showed that it is possible to either increase or decrease the regularity of spontaneous spike trains by blocking some portion of either potassium or sodium ion channels. Recently, Gong et al. \cite{r16} extended the subject by examining the effects of channel blockage on the collective spontaneous spiking activity of coupled stochastic HH neurons, and found that the latter can be reduced or enhanced by blocking the sodium or potassium channels within small or large cell sizes. They constructed the interaction network as an array of bi-directionally coupled neurons. However, small-world (SW) networks, combining high clustering with scattered long range connections \cite{r17} have been suggested as an attractive option to search for connectivity information of both anatomical and functional networks in the brain, mainly because this topology can support both local and distributed information processing \cite{r18}. Therefore, SW networks have been widely used to understand how neuronal circuitry generates complex patterns of activity \cite{r19,r20,r21,r22,r23}.

Our aim in this Letter is to extend the subject by using SW networks as the underlying interaction topology between neurons, and to investigate the impact of sodium and potassium channel blockage on the spontaneous collective spiking regularity as a function of the network structure and cell size in a manner that more closely mimics actual conditions.

\section {Mathematical model and setup}
Within this study we use the HH model of neuronal dynamics as the basic building block of the investigated system, whereby the time evolution of the membrane potential for coupled neurons on an arbitrary network in the absence of deterministic external stimuli is given as follows:
{\setlength\arraycolsep{0.1em}\begin{eqnarray}
C_{m}{\rm d}V_{i}/{\rm d}t=&&G_{Na}(m_i,h_i)(V_{Na}-V_i)+G_K(n_i)(V_K-V_i)\nonumber\\*
&&+G_L(V_L-V_i)+\sum_{j}\varepsilon_{ij}(V_j-V_i)
\end{eqnarray}}
Here $V_i$ denotes the membrane potential of neuron $i=1,\ldots,N$ ($N$ being the system size), $C_m=1{\rm\mu Fcm^{-2}}$ is the membrane capacity, and $G_{Na}$, $G_K$ and $G_L$ represent sodium, potassium and leakage conductances, respectively. $V_{Na}=50{\rm mV}$, $V_{K}=-77{\rm mV}$ and $V_{L}=-54.4{\rm mV}$ are the reversal potentials for the sodium, potassium and leakage channels, respectively. Moreover, $\varepsilon_{ij}$ denotes the coupling strength between neurons $i$ and $j$, whereby we set $\varepsilon_{ij}=\varepsilon$ if the two are connected or $\varepsilon_{ij}=0$ otherwise. In the model, the leakage conductance is assumed to be constant, equaling $G_{L}=0.3{\rm mScm^{-2}}$, while the sodium and potassium conductances change dynamically according to the following two equations:
\begin{equation}
G_{Na}(m_i,h_i)=g^{max}_{Na}x_{Na}m^{3}_{i}h_{i},\
G_{K}(n_i)=g^{max}_{K}x_{K}n^{4}_{i}
\end{equation}
Here $g^{max}_{Na}=120{\rm mScm^{-2}}$ and $g^{max}_{K}=36{\rm mScm^{-2}}$ are the maximal sodium and potassium conductances, respectively. Moreover, $m$ and $h$ denote the activation and inactivation gating variables for the sodium channel, respectively, whereas the potassium channel includes an activation gating variable $n$. In Eq. (2) we also introduce two scaling factors, $x_{Na}$ and $x_K$, which are the fractions of non-blocked ion channels to the total number of sodium ($N_{Na}$) or potassium ($N_K$) ion channels within the patch area, respectively \cite{r14,r15,r16}. These factors are confined to the unit interval.

In the HH model, activation and inactivation gating variables, $m_i$, $n_i$ and $h_i$, change over time in response to the membrane potential following first-order differential equations within the limit of very large cell sizes. However, when the population of ion channels is finite, the stochastic behavior of voltage-gated ion channels (or ion channel noise) must be taken appropriately into account. In this study, we follow the approach in previous works related to the channel block \cite{r14,r15,r16}, and use the algorithm proposed by Fox \cite{r25}. In the Fox's algorithm, variables of stochastic gating dynamics are described with the corresponding Langevin generalization
\begin{equation}
{\rm d}x_{i}/{\rm d}t=\alpha_x(1-x_i)-\beta_x x_i+\xi_{x_{i}}(t),\
x_i=m_i,n_i,h_i
\end{equation}
where $\alpha_x$ and $\beta_x$ are rate functions for the gating variable $x_i$. The probabilistic nature of the channels appears as a source of noise $\xi_{x_{i}}(t)$ in Eq. (3), which is an independent zero mean Gaussian noise whose autocorrelation function is given as follows \cite{r25}:
\begin{equation}
\langle \xi_m(t) \xi_m(t') \rangle = \frac{2 \alpha_m \beta_m}{N_{Na}x_{Na}(\alpha_m + \beta_m)} \delta(t-t')
\end{equation}
\begin{equation}
\langle \xi_h(t) \xi_h(t') \rangle = \frac{2 \alpha_h \beta_h}{N_{Na}x_{Na}(\alpha_h + \beta_h)} \delta(t-t')
\end{equation}
\begin{equation}
\langle \xi_n(t) \xi_n(t') \rangle = \frac{2 \alpha_n \beta_n}{N_{K}x_{K}(\alpha_n + \beta_n)} \delta(t-t')
\end{equation}
where the factors, $x_{Na}$ and $x_K$, are used again to disregard the blocked channels, which do not contribute to the intrinsic channel noise. Given the assumption of homogeneous sodium and potassium ion channel densities, channel numbers are calculated via $N_{Na}=\rho_{Na}S$, $N_{K}=\rho_{K}S$ where $\rho_{Na}=60{\rm\mu m^{-2}}$ and $\rho_{K}=18{\rm\mu m^{-2}}$ are the sodium and potassium channel densities, respectively, whereas $S$ represents the total membrane area or the cell size \cite{r25}. Equations (1)-(6) constitute the stochastic HH model, where the cell size $S$ determines the intrinsic noise level through the number of ion channels $N_{Na}$ and $N_{K}$. When $S$ is large stochastic effects related to the channel noise are negligible due to large numbers of ion channels in Eqs. (4)-(6), and thus the intrinsic channel noise appearing in Eq. (3) for the gating variables vanishes. Accordingly, the stochastic model then approaches the deterministic description. However, when the number of ion channels (or the cell size $S$) is small, stochastic effects have a significant impact on the membrane dynamics \cite{r14,r15,r16,r25,add5}.

\begin{figure}
\center \scalebox{0.4}[0.4]{\includegraphics{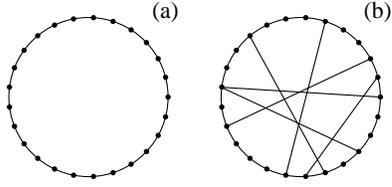}}
\caption{Examples of considered network topologies. For clarity regarding $k$ and $p$ only $N=25$ vertices are displayed in each panel. (a) Regular ring characterized by $p=0$ with periodic boundary conditions. Each vertex is connected to its $k=2$ nearest neighbors. (b) Realization of Newman-Watts small-world topology via randomly adding $M$ new links (in this case 6 new links were added, hence $p=0.02$).}
\end{figure}

Furthermore, the interaction network is comprised of identical HH neurons, initially each having connectivity $k=2$, with the system size set to $N=60$. Following the Newman-Watts model \cite{r26}, we start with a regular ring [see Fig. 1(a)] and make a random draw of two neurons. Subsequently, if they are not already connected, we add a non-directed link between them. This process is repeated until a total of $M$ new links are added [see Fig. 1(b)], finally resulting in a network which is equivalent to the one obtained if new edges would be added with probability $p=2M/[N(N-1)]$.

In order to asses the system's dynamics, we measured the collective temporal behavior of the network by first calculating the average membrane potential, $V_{avg}(t)=N^{-1}\sum^{N}_{i=1} V_i(t)$, corresponding to the mean field of the network, and then quantifying the spontaneous collective spiking regularity by the coefficient of variation ($CV$) of the inter-spike intervals ($ISI$s) according to:
\begin{equation}
CV=\frac{\sigma_{ISI}}{\langle ISI \rangle}=
\frac{\sqrt{\langle ISI^2 \rangle - \langle ISI \rangle^2}}{\langle ISI \rangle}
\end{equation}
Here `spike' times were defined by the upward crossing of $V_{avg}(t)$ past a detection threshold of 0mV  (referred to below as `network spikes'), whereas $\langle ISI \rangle$ and $\langle ISI^2 \rangle$  denote the mean and the mean squared inter-spike intervals, respectively. Notably, $CV$ is characterized by smaller values for more ordered spike trains and vanishes for a deterministic signal. In the results presented below, for any given set of the remaining network parameters (scaling factor, noise level, coupling strength), quantitative results re pooled from simulations of 10 realizations of the network for any given value of $p$.

\section {Results and discussion}

\begin{figure}
\center \scalebox{0.48}[0.48]{\includegraphics{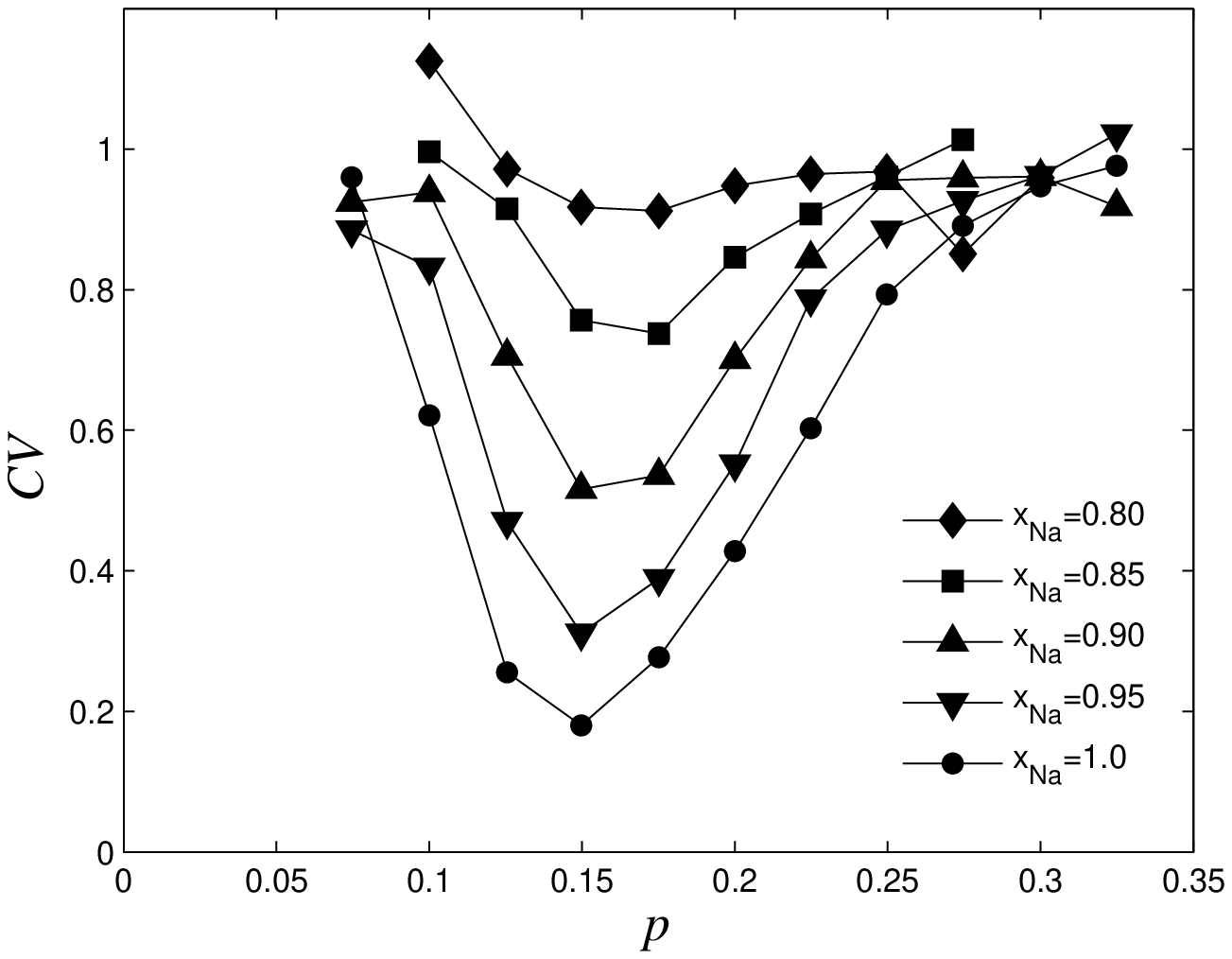}}
\center \scalebox{0.48}[0.48]{\includegraphics{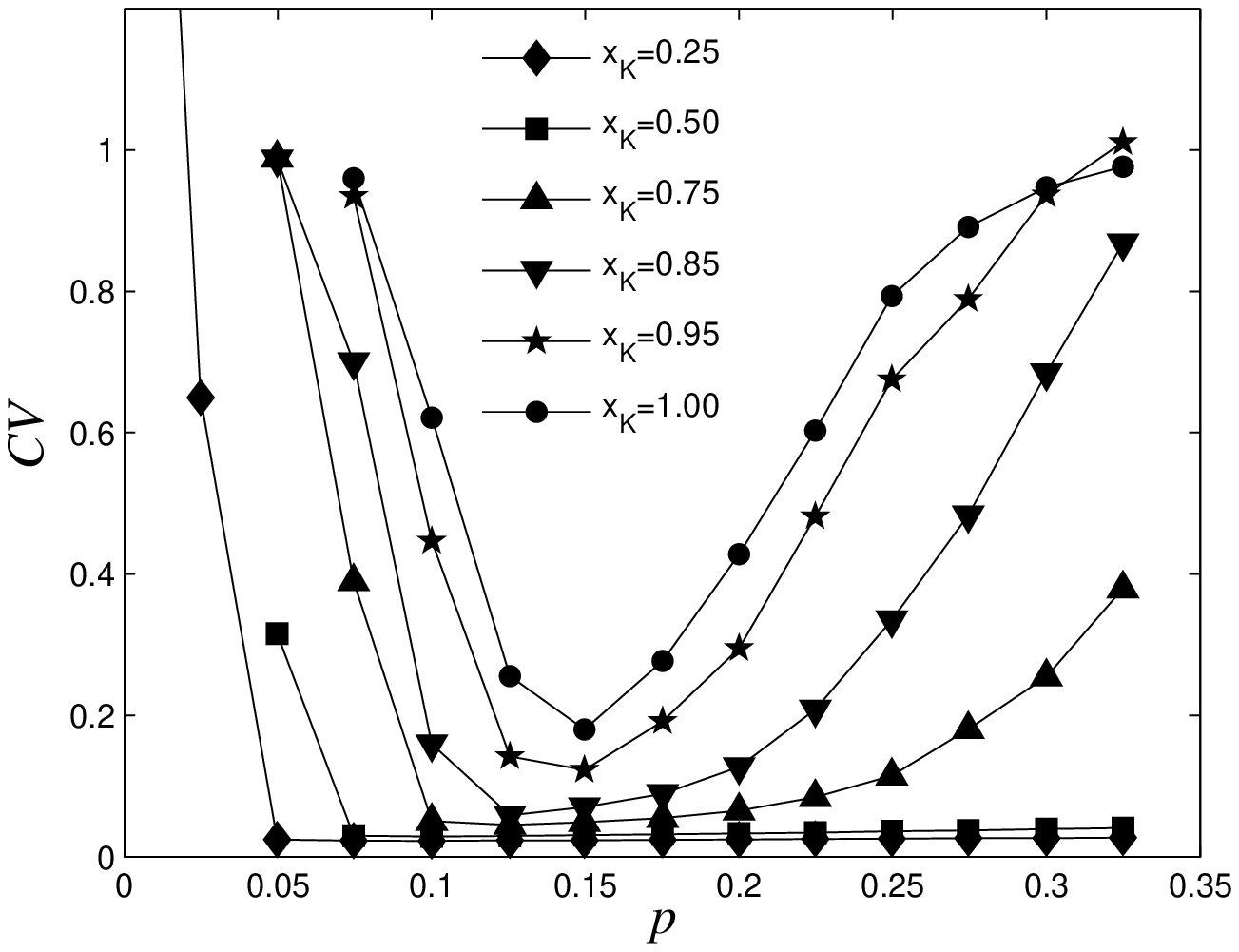}}
\caption{The dependence of the spontaneous spiking regularity ($CV$) on $p$ obtained by a fixed patch area $S=6{\rm\mu m^{2}}$ and coupling strength $\varepsilon=0.1$. (a) Different levels of sodium channel block. (b) Different levels of potassium channel block.}
\end{figure}

In what follows, we systemically analyze the impact of sodium and potassium ion channel block on the spontaneous collective spiking regularity of the network as a function of the network topology and cell size (thus, ion channel noise). First, we investigate how $CV$ changes with the scaling factors, $x_{Na}$ or $x_K$, as a function of the network topology by a fixed cell size $S=6{\rm\mu m^{2}}$ and coupling strength $\varepsilon=0.1$. Thereby we vary the density of one channel type (either $x_{Na}$ or $x_K$) while keeping the other equal to one. Obtained results are presented in Fig. 2(a) for sodium and in Fig. 2(b) for potassium ion channel block. Notably, we did not consider $x_{Na}<0.8$ because then the average membrane potential $V_{avg}$ did not include spikes. For all values of $x_{Na}$ and for all $x_{K}>0.5$ considered in Fig. 2, the $CV$ exhibits a well-expressed minimum by an optimal fraction of random shortcuts $p$, at which the spontaneous collective spiking is most regular. Moreover, it is evident that there exists an optimal $p$ independent of the scaling factors, thus indicating the robust existence of topology-dependent coherence resonance \cite{r27,r28,r29,r30}. Interestingly, an optimal topology is obtained at $p \approx 0.15$ for both the sodium as well as the potassium channel blocks, indicating the existence of a universally optimal network structure. However, as $x_K$ decreases to $x_K \le 0.5$, the minimal values of $CV$ are obtained already by slightly lower values of $p$, yet after reaching the minimum, the regularity of $V_{avg}$ becomes constant and virtually independent of the network topology. Indeed, the level of $x_K$ in the range $\le 0.5$ plays a rather insignificant role for the temporal regularity of the average membrane potential for all $p \ge 0.075$, as can be inferred from the bottom two curves depicted in Fig. 2(b).

Furthermore, it is worth noting that the reduction of working sodium channels [decrease of $x_{Na}$ in Fig. 2(a)] decreases the collective spiking regularity, whereas a reduction of working potassium channels [decrease of $x_{K}$ in Fig. 2(b)] increases it by a fixed cell size, as reported also in \cite{r14,r15} for a single neuron. Notably, Gong \textit{et al.} \cite{r16} found that sodium or potassium channel blocks can either enhance or reduce the collective spiking regularity of an array of bi-directionally coupled neurons, depending on the cell size. In our case, however, for $p \approx 0.15$ the collective regularity is very high for all values of $x_K$ and $x_{Na}>0.9$ if compared to the values reported in \cite{r16} for the same cell size and scaling factors (see Figs. 1 and 2 in \cite{r16}). This indeed strongly supports the fact that the fine-tuning of the network structure via $p$ is able to significantly facilitate the spontaneous spiking regularity of neurons in the presence of either sodium or potassium ion channel block.

\begin{figure}
\center \scalebox{0.48}[0.48]{\includegraphics{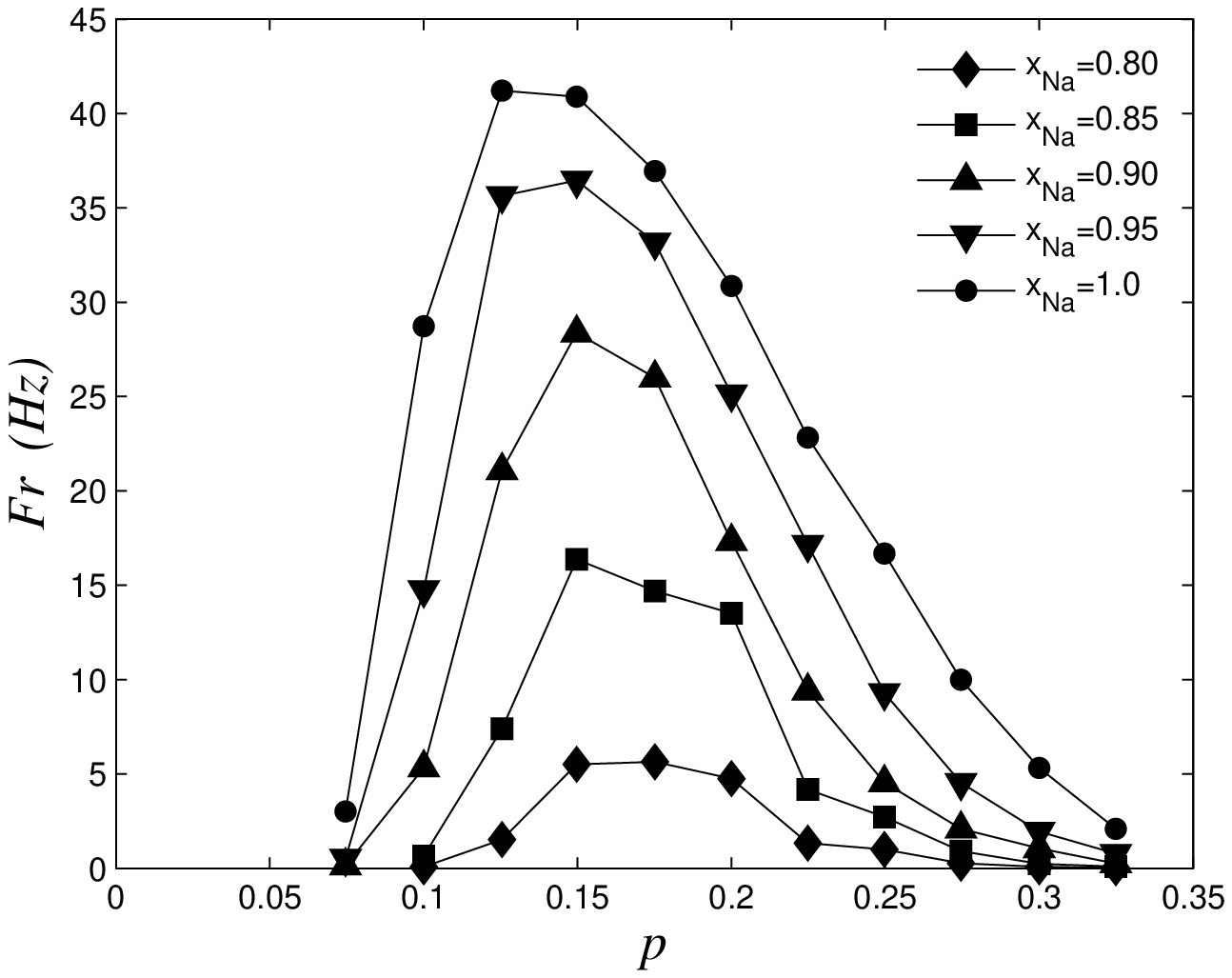}}
\center \scalebox{0.48}[0.48]{\includegraphics{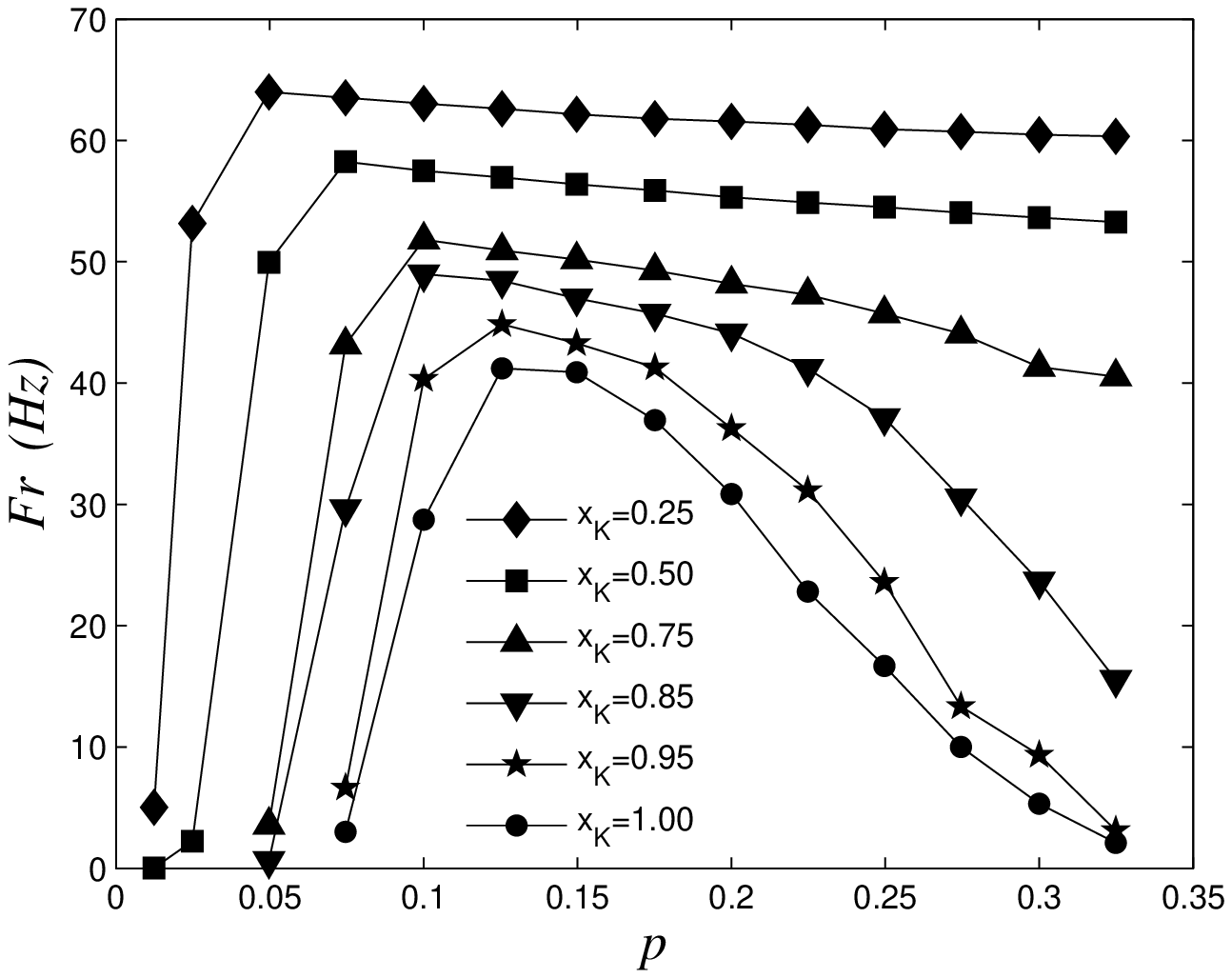}}
\caption{The dependence of the mean firing rate ($Fr$) on $p$ obtained by a fixed patch area $S=6{\rm\mu m^{2}}$ and coupling strength $\varepsilon=0.1$. (a) Different levels of sodium channel block. (b) Different levels of potassium channel block.}
\end{figure}

To gain more insight into the dependence of the collective spiking regularity on $p$  and the scaling factors, we further calculate the mean firing rate ($Fr$) for different values of $p$ and $x_{Na}$ or $x_K$. Obtained results are presented in Fig. 3(a) for sodium and in Fig. 3(b) for potassium ion channel block. In case of sodium channel block, the mean firing rate exhibits a resonance-like behavior with a distinct maximum appearing by the optimal fraction of random shortcuts equaling approximately $p \approx 0.15$. Potassium channel block leads to similar behavior, but only when $x_K \to 1$. As $x_K$ decreases below 0.95, the maxima are achieved by fewer added shortcuts (lower values of $p$). Thus, the outlay of $CV$ presented in Fig. 2 follows the characteristics of the firing rate, and moreover, decreases with higher $Fr$ (compare Figs. 2 and 3). In this sense, results presented in Fig. 3 are consistent with those presented in Fig. 2 for both types of ion channel blockage. In particular, the refractory period of spike generation postulates a certain minimal time that needs to elapse between two spikes. Thus, as the firing frequency increases, the mean $ISI$ decreases and can approach the refractory period \cite{r15}. In other words, as the firing frequency increases, the refractory period becomes more significant relative to the diminishing $ISI$, contributing to the reduction of the output $CV$ \cite{r31,r32}. We therefore argue that the network topology affects the regularity of spontaneous neuronal spiking by means of serving as a scaling factor for the firing rate.

Related to the above argumentation, it is interesting to elaborate further on the results obtained for $x_K \le 0.5$. There a further increase in $p$, past the optimal value for which the maximal firing rate is obtained, decreases the mean firing rate fairly slightly [see the upper two curves in Fig. 3(b)]. As shown in the bifurcation diagram for the deterministic autonomous HH model in \cite{r15} [see their Fig. 1(b)], with decreasing the potassium conductance via $x_K$ in Eq. (2), a sub-critical Hopf bifurcation occurs at $x_K=0.549$, after which the system enters a region of stable oscillatory spiking solutions within the purely deterministic description \cite{r14,r15}. We calculated the firing rate of the deterministic network by considering just the conductance reductions $x_K = 0.5$ and $x_K = 0.25$, and found them equaling 51Hz and 61Hz, respectively. These values are almost equal to those obtained for the model with stochastic components as $p$ approaches to its upper boundary (all-to-all connected network) [see Fig. 3(b)]. This suggests that the stochastic dynamics has little effect on the spontaneous regularity for the potassium channel block if $x_K \le 0.5$. It also suggests that the network connectivity has a negligible impact on the firing rate (or $CV$) for all $x_K \le 0.5$ as $p \to 1$. Importantly however, for small values of $p$ the firing rate reaches a maximum above its deterministic value, indicating the existence of an optimal small-world topology that warrants the best regularity of spontaneous collective spiking activity even in a region with a purely deterministic description, which then again disappears as $p \to 0$ (regular nearest-neighbor network).

\begin{figure}
\center \scalebox{0.48}[0.48]{\includegraphics{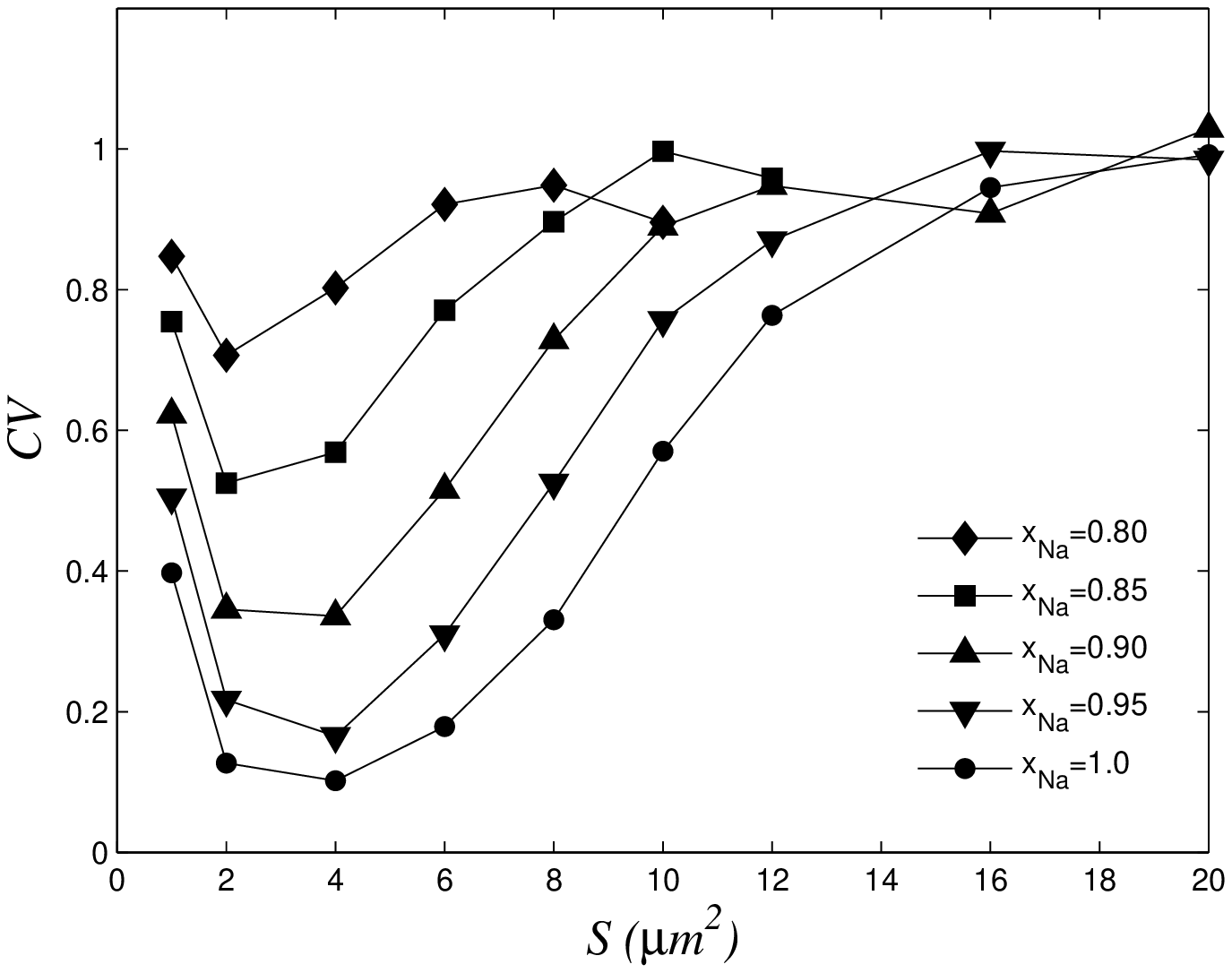}}
\center \scalebox{0.48}[0.48]{\includegraphics{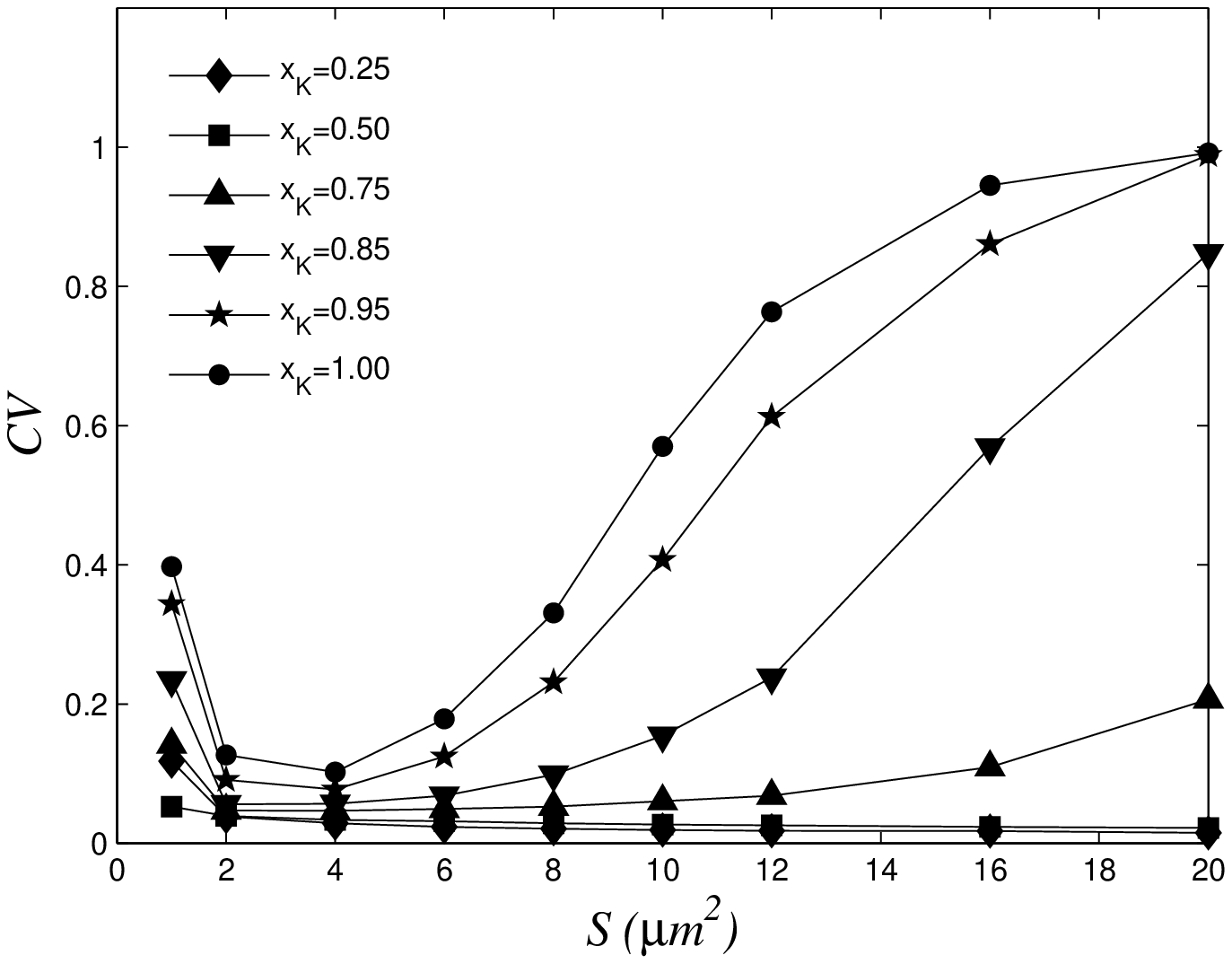}}
\caption{The dependence of the spontaneous spiking regularity ($CV$) on the patch area $S$ obtained by the coupling strength $\varepsilon=0.1$ and $p=0.15$. (a) Different levels of sodium channel block. (b) Different levels of potassium channel block.}
\end{figure}

Following above investigations, where we found that the CV exhibits a well-expressed minimum at an optimal $p \approx 0.15$ by a fixed cell size $S=6{\rm\mu m^{2}}$, we investigate now how the regularity by the optimal value of $p$ changes with the cell size for different values of the scaling factors determining the non-blocked channel densities $x_{Na}$ and $x_K$. Obtained results are presented in Fig. 4(a) for sodium and in Fig. 4(b) for potassium ion channel block. We find that the spontaneous collective spiking regularity depends strongly on the cell size for all values of $x_{Na}$ and for all $x_{K}>0.5$, with the best regularity emerging for small, but non-zero cell sizes of around $S=2-4{\rm\mu m^{2}}$ constituting a specific intensity of intrinsic noise. For this range of $S$ values, the regularity changes slightly depending on $x_{K}$, whereas it may change rather substantially depending on $x_{Na}$. However, for larger cells (large $S$) the $CV$ differs slightly depending on $x_{Na}$ while it differs substantially depending on $x_K$.

On the other hand, for $x_K \le 0.5$ an increase in the cell size beyond a value of around $S=2{\rm\mu m^{2}}$ almost does not improve the regularity of the output, albeit a small improvement in $CV$ can be inferred for $x_K = 0.25$ if compared to $x_K = 0.5$. This collective behavior seems similar to that of a single HH neuron reported in \cite{r14,r15}, where the spontaneous regularity was found to increase as the cell size increased for $x_K \le 0.5$. However, we show here that this increase is larger for a single neuron if compared to that of the whole network. Also, we note that the behavior of an array of bi-directionally coupled neurons reported in \cite{r16} was found to be rather different, where indeed the regularity decreased as the cell size increased past $S \ge 2{\rm\mu m^{2}}$.

\begin{figure}
\center \scalebox{0.48}[0.48]{\includegraphics{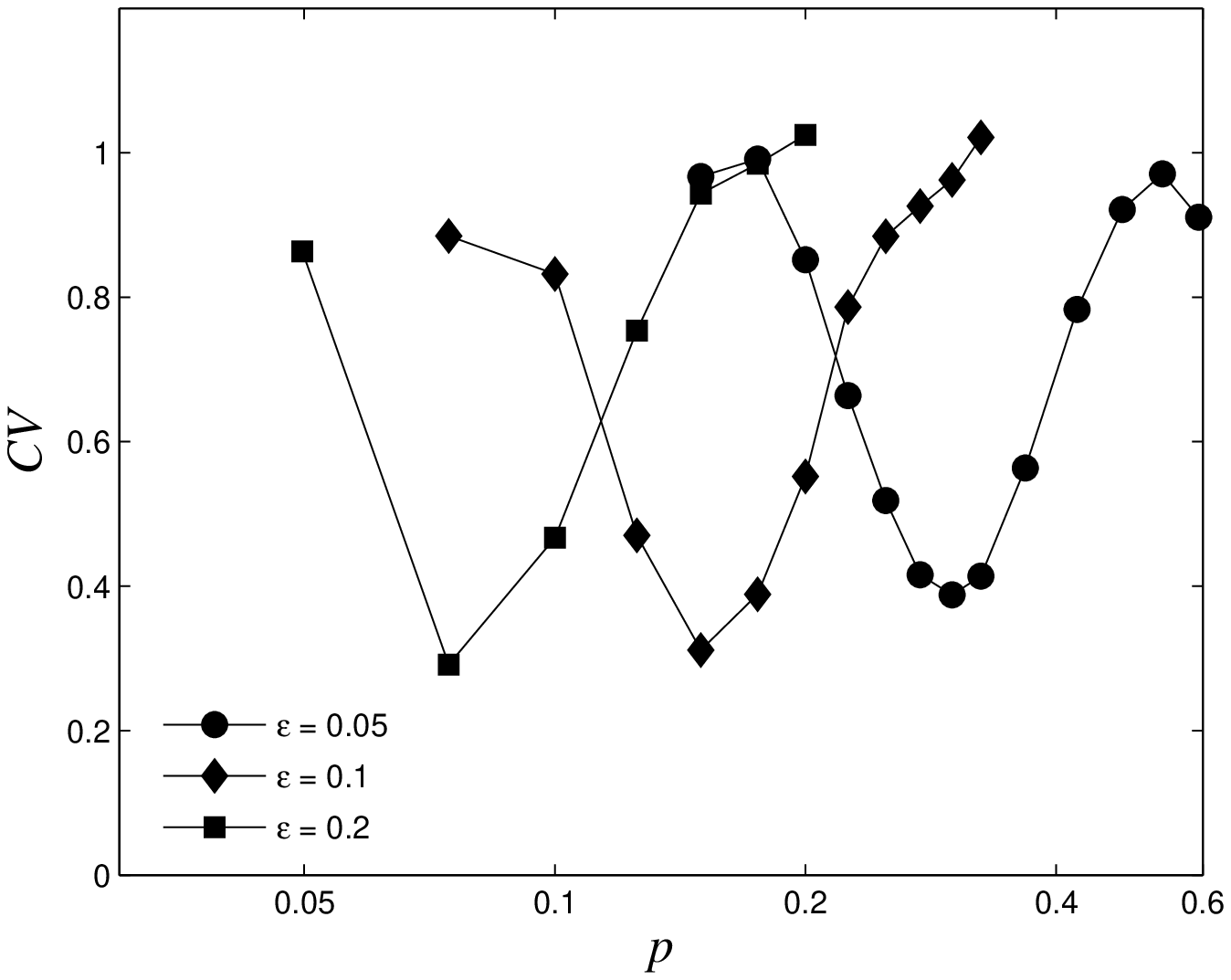}}
\center \scalebox{0.48}[0.48]{\includegraphics{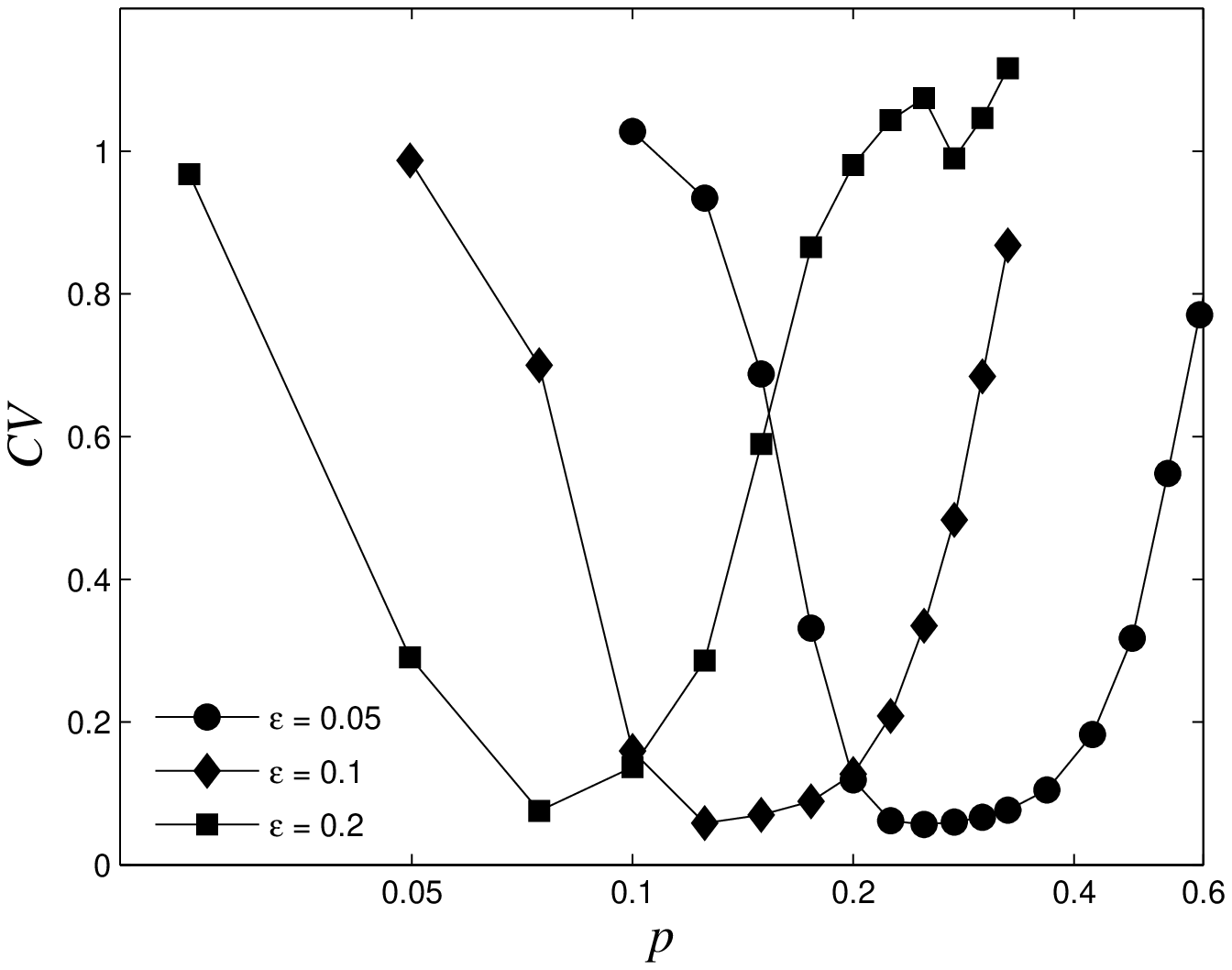}}
\caption{The dependence of the spontaneous spiking regularity ($CV$) on $p$ for three different coupling strengths $\varepsilon$ obtained by a fixed patch area $S=6{\rm\mu m^{2}}$. (a) Sodium channel block is $x_{Na}=0.95$. (b) Potassium channel block is $x_{K}=0.85$.}
\end{figure}

Notably, we also investigated the $CV$ dependencies presented in Fig. 4 via the mean firing rate (not shown), and arrived at the findings that are in agreement with the above-established link between the $ISI$ and the firing rate when interpreting results presented in Figs. 2 and 3. The latter support the fact that, not only does the network topology affect the regularity of spontaneous neuronal spiking by means of serving as a scaling factor for the firing rate, but the regularity by the optimal SW topology is affected also by the intensity of intrinsic noise, similarly as reported earlier for the internal-noise coherence resonance. These two facts combined yield a doubly coherence resonance-like phenomenon \cite{r33,r34} in our case, related to the regularity of spontaneous neuronal spiking, that is brought about by the fine-tuning of both $p$ and $S$.

In order to further extend and conclude the study, we investigate the dependence of the spontaneous collective regularity on $p$ for three different coupling strengths $\varepsilon$, by a fixed cell size $S=6{\rm\mu m^{2}}$, and two different cases of ion channel block $x_{Na}=0.95$ and $x_{K}=0.85$. The two values were selected so as to lead to a coherence resonance in $CV$ against $p$. Obtained results are presented in Fig. 5. From them we arrive at two findings that are independent of the type of the ion channel block. First, the resonance-like outlay of $CV$ in dependence on $p$ is similar for all coupling strengths and its overall minimum does not change. Second, the coupling strength is inversely correlated with the value of $p$ that yields the minimum $CV$. Accordingly, as $\varepsilon$ increases the minimum $CV$ is obtained by fewer and fewer added shortcuts (by lower values of $p$). However, these correlations are not completely valid for the total span of $p$, because irrespective of $\varepsilon$, the spontaneous collective regularity becomes very low for extreme values of $p$ (\textit{i.e.} $p \to 0$ or $p \to 1$). Nevertheless, the findings are consistent with previous results related to different subjects but to the same network structure \cite{r23,r35}.

\section {Summary}
In sum, we show that the regularity of spontaneous spiking activity can be resonantly enhanced via fine-tuning of both $p$ and $S$. The extend of improved regularity is thereby maximal by high $x_{Na}$ and low $x_{K}$ values. Solely by potassium ion channel blockage constituted by $x_K \le 0.5$ the resonant dependence on $p$ vanishes and the first occurrence of the minimal $CV$ occurs already by somewhat lower values of $p$. There also exists an optimal intensity of internal noise, constituted by cell sizes spanning $S=2-4{\rm\mu m^{2}}$, by which the regularity of spontaneous spiking activity on small-world networks is maximal. Such cell sizes may correspond to so-called `hot spots' \cite{add1} or otherwise spatially restricted regions in dendritic or axonal trees \cite{add2,add3}. Both resonant-like dependencies by different values of $x_{Na}$ and $x_{K}$ can be nicely corroborated by the analysis of the firing rate as well as the deterministic model dynamics. Finally, we show that the resonant outlay of the $CV$ dependence on $p$ prevails irrespective of the coupling strength, although the minima shift in accordance with some previous findings. Although the present model considers the intrinsic noise stemming from the stochastic dynamics of ion channels, our study also provides some indication towards the potential impact of synaptic noise because the latter may be approximated by a non-zero mean Gaussian white noise \cite{add4}. We hope our study will prove useful in striving towards the understanding of importance of structure \cite{r36,r37,r38,r39} and uncertainty in active neuronal networks.

\end{document}